\documentclass[12pt,reqno]{amsart}
\usepackage[cp1251]{inputenc}
\usepackage[english,russian]{babel}
\usepackage{amsmath,amssymb,amsfonts,cite}

\numberwithin{equation}{section}

\let\p\partial

\let\cal\mathcal

\def\phi{\varphi}

\let\ge\geqslant
\let\le\leqslant

\def\ba{\begin{aligned}} 
\def\ea{\end{aligned}}

\newcommand{\pa}{\partial}

\newcounter{rem}

\numberwithin{equation}{section}
 
\begin{document}
\baselineskip=6.5mm
\thispagestyle{empty}

\baselineskip=7mm
\thispagestyle{empty}
\begin{center}
{\Large\bf Algebraic quantum Hamiltonians on the plane}
\end{center}

 \vskip5mm \hfill
\begin{minipage}{12.5cm}
\baselineskip=15pt
{\qquad \qquad \bf V.V. Sokolov ${}^{1}$} \\ [1ex]
{\footnotesize
${}^1$ Landau Institute for Theoretical Physics, Moscow, Russia
 }\\

\end{minipage}
\begin{center}
\begin{minipage}[c]{140mm}
\small ABSTRACT.  We consider second order differential  operators $P$ with polynomial coefficients that preserve the vector space 
$V_k$ of polynomials of degrees not greater then $k$. We assume that the metric associated with the symbol of $P$ is flat and that the operator $P$ is potential. In the case of two independent variables we obtain some classification results and find polynomial forms for   the elliptic $A_2$ and $G_2$ Calogero-Moser Hamiltonians and for the elliptic Inosemtsev model. 
\end{minipage}
\end{center}

\section{\bf \large Introduction}

\bigskip

\qquad In the paper {\cite{sokturb}} a transformation that brings the elliptic Calogero-Moser operator
\begin{equation}\label{CM3}
H=\Delta+ g \sum_{i>j}^3 \wp(x_i-x_j), 
\end{equation}
where $g$ is arbitrary constant, to a differential operator with polynomial coefficients.  
This problem is easily reduced  to two-dimensional case. Indeed, in variables  $y_i=x_i-\frac{Y}{3},$ where $Y=\sum_{i=1}^3 x_i$ 
the Laplacian is given by 
$$
\Delta=-3 \frac{\partial^2}{\partial Y^2}-\frac{2}{3}\Big(\frac{\partial^2}{\partial y_1^2}+\frac{\partial^2}{\partial y_2^2}-\frac{\partial^2}{\partial y_1 \partial y_2} \Big).
$$
The variable $Y$ is separated and we reduce (\ref{CM3}) to 
\begin{equation}\label{HAM}
 {\cal H}= -\frac{1}{3}\Big(\frac{\partial^2}{\partial y_1^2}+\frac{\partial^2}{\partial y_2^2}-\frac{\partial^2}{\partial y_1 \partial y_2} \Big)+\bar g \sum_{i>j}^3 \wp(y_i-y_j), 
\end{equation}
where $y_3=-y_1-y_2.$ It turns out that the change of variables 
\begin{equation} \label{ellip}
x=\frac{\wp'(y_1)-\wp'(y_2)}{\wp(y_1) \wp'(y_2)-\wp(y_2) \wp'(y_1)}, \qquad y=\frac{\wp(y_1)- \wp(y_2)}{\wp(y_1) \wp'(y_2)-\wp(y_2) \wp'(y_1)}
\end{equation}
and a gauge transformation of the form $L\to f L f^{-1},$ where $f$ is an appropriate function in $x,y, $
bring (\ref{HAM}) to a differential operator of the form 
\begin{equation}
\label{oper}
  L\ =\ a(x,y) \frac{\pa^2}{\pa x^2}+2 b(x,y) \frac{\pa^2}{\pa x \pa y}+c(x,y) \frac{\pa^2}{\pa y^2}
 + d(x,y) \frac{\pa}{\pa x}\ + e(x,y) \frac{\pa}{\pa y}+f(x,y)
\end{equation}
with polynomial coefficients. We call $L$ {\it polynomial form} of (\ref{HAM}). The set of coefficients $a,b,$ and $c$ is called 
the {\it symbol} of $L$.

A polynomial form for the elliptic case have been found in  {\cite{sokturb}} by a deformation of known \cite{RT:1995} 
polynomial form of the trigonometric model. Concerning polynomial forms for the elliptic Calogero-Moser see \cite{matush}.

The goal of this paper is a description and an investigation of a class of operators of the form (\ref{oper}) that contains polynomial forms o many operators related to simple Lie algebras  \cite{perel}. This class is selected by postulating of some properties of polynomial form for operator (\ref{HAM}).

It is clear that for this polynomal form $L$  

{\bf 1:} the contravariant metric $g$ defining by the symbol of $L$ is flat $\square$
\newline and 

 {\bf 2:} $L$ can be reduced to a self-adjoint operator by a gauge transformation $L\to f L f^{-1}.$   $\square$

Furthermore,  it can be shown that for some special values of the parameter $\bar g$ in (\ref{HAM}) 

 {\bf 3:} the operator $L$ preserves the vector space $V_n$ of all polynomials $P(x,y)$ such that ${\rm deg}\, P\le n$ for some $n\ge 2. $ $\square$
 
and that

{\bf 4:} $L$ is invariant with respect to the involution $\tilde x=x,\, \tilde y=-y.$ $\square$
 
In Section 2 we consider diffrential operators of second order with one independent variable. In this case the assumptions {\bf 1-2} are automatically satisfied. We describe all operators that satisfy Property {\bf 3}. As a result we arrive at the polynomial forms for the Lame operator and for the more general  Darboux operator \cite{Darbu}.

Our main goal is to investigate two-dimensional polynomial operators (\ref{oper}) that satisfy Properties {\bf 1-3}. 
In Section 3.1 we describe operators that satisfy Property {\bf 3}, define an action of the group  $SL_3$ on the vector space of such operators, and find some of invariants for this action. 

In Section 3.2 we find all operators with elliptic and trigonometric symbols (see Definition 1) that satisfy Properties  {\bf 1-4}.     As a result, we arrive at  polynomials forms for the elliptic $A_2$ and $G_2$-Hamiltonians and for the Inozemtsev elliptic model. The $A_2$ and $G_2$ polynomial forms are equivalent to ones presented in \cite{}. The polynomial form for the Inozemtsev model is probably new. Possibly it can be deduce from the rational form found in \cite{Takemura}. It would be interesting to find the Hamiltonians in the  Schrodinger form corresponding to five trigonometric sysmbols obtained during the classification. 
 
By miracle, all known operators $L$ satisfying Properties 1-3 turn out to be integrable (i.e. possess differential operators that commute with L). 

It is clear that if a differential operator obeys Property 3, we can find several eigenvalues and corresponding polynomial eigenvectors by means of linear algebra. In \cite{Turbiner:1988}  such operators are called {\it quasi-exact solvable}. 

For the elliptic models the flat metric $g$ depends on the elliptic parameter. One of the reasons why such a metric could be interesting in itself is  that families of contravariant metrics with linear dependence on a parameter are closely related to the  Frobenius manifolds  \cite{dub}

\vskip.3cm
\noindent
{\bf Acknowledgments.}
The author is grateful to M. Matushko, A. Turbiner, and E. Ferapontov for useful discussions. He is also grateful to  MPIM(Bonn)
for hospitality and financial support. 

\section{ \bf \large Differential operators with one independent variable}

Suppose that an  order $m$ polynomial operator of the form  
 \begin{equation} \label{ODEoper}
Q=\sum_{i=0}^m  a_i(x)\frac{ d^i}{dx^i}
\end{equation} 
preserves  the vector space $V_n$ of all polynomials of degree $\le n$, where $n\ge m$.  

{\bf Conjecture} (A. Turbiner \cite{Turbiner:1994}). Any such operator $Q$ 
 is  a (non-commutative) polynomial of order $m$ in first order generators
$$
J_1=1, \qquad J_2 = \frac{d}{dx}, \qquad J_3 = x \frac{d}{dx}, \qquad J_4 = x^2 \frac{d}{dx}- n x. \qquad \square
$$
Notice that the Lie algebra generated by $J_1,...,J_4$ is isomorphic to $gl(2).$

For small $m$ this statement can be straightforwardly verified using the following 

{\bf Lemma 1.} If operator (\ref{ODEoper}) preserves the vector space $V_n$, where $n\ge m,$ then the degree of the polynomial $a_i$ is not greater then $m+i$.

{\bf Proof.} Define a weighting scheme in the ring of differential operators with polynomial coefficients by setting weight$(x)= 1$ and  weight $\frac{\pa}{\pa x}=-1$. Let
$
Q=\sum_{i=-m}^p Q_i
$
be the decomposition of $Q$ into the sum of homogeneous differential operators $Q_i$, where weight$(Q_i)=i$. We have to show that $p\le m.$ Indeed, if $p>m$ then $Q_p(x^n)=Q_p(x^{n-1})=\cdots =Q_p(x^{n-m})=0.$ Since the kernel of the differential operator $Q_p$ of order $m$ contains $m+1$ linearly independent functions, we get $Q_p\equiv 0.$ $\square$

 It  follows from Lemma 1 that any operator $P$ of second order that preserves $V_n$ has the following structure:
\begin{equation}\label{P11}
P= (a_4 x^4+a_3 x^3+a_2 x^2+a_1 x +a_0)  \frac{ d^2}{dx ^2} +(b_3 x^3+b_2 x^2+b_1 x +b_0)  \frac{ d}{dx}+c_2 x^2+c_1 x+c_0. 
\end{equation}
Taking into account that $P(x^n)$ and $P(x^{n-1})$ have to be polynomials of degree not greater then $n,$ we find that the constants in the coefficients of (\ref{P11}) are connected by the following identies
\begin{equation}\label{P22}
b_3=2 (1-n)\, a_4 ,\qquad c_2= n (n-1) \, a_4, \qquad c_1=n (a_3-n a_3-b_2). 
\end{equation}

The transformation group
\begin{equation}\label{flin}
x\to \frac{s_1 x+s_2}{s_3 x+s_4}, \qquad P\to  (s_3 x+s_4)^{-n} P (s_3 x+s_4)^n,
\end{equation}
acts on the nine-dimensional vector space of such operators. The coefficient $a(x)$  at the second derivative is a fourth order polynomial, which transforms as follows
\begin{equation}\label{transa}
a(x)\to (s_3 x+s_4)^4 a\Big(\frac{c_1 x+c_2}{c_3 x+c_4}\Big).
\end{equation}
If $a(x)$ has four distinct roots on the Riemann sphere, we call the operator $P$ {\it elliptic}. In the elliptic case using transformations (\ref{transa}), 
we may reduce $a$ to
$$
a(x)=4\, x(x-1)(x-\kappa),
$$ 
where $\kappa\ne 0,1$ is the elliptic parameter. 
 
Define parameters $n_1,...,n_5$ by identities
$$
b_0=2 (1+2 n_1), \quad b_1=-4 \Big((\kappa+1)(n_1+1)+\kappa n_2+n_3 \Big),\qquad b_2=-2\,(3+2 n_1+2 n_2+2 n_3),
$$
$$
n=- \frac{1}{2}(n_1+n_2+n_3+n_4), \quad n_5=c_0+ n_2(1-n_2)+\kappa n_3 (1-n_3)+(n_1+n_3)^2+\kappa (n_1+n_2)^2.
$$
Then the operator $H=h P h^{-1},$ where 
$$\displaystyle  h=x^{\frac{n_{1}}{2}}(x-1)^{\frac{n_{2}}{2}}(x-\kappa)^{\frac{n_{3}}{2}}$$
has the form
$$
a(x) \frac{ d^2}{dx ^2}+\frac{a'(x)}{2} \frac{ d}{dx}+n_5+n_4 (1-n_4)\,x+
\frac{n_1 (1-n_1) \kappa}{x}+\frac{n_2 (1-n_2) (1-\kappa)}{x-1}+\frac{n_3 (1-n_3) \kappa (\kappa-1)}{x-\kappa}.
$$
After the transformation $x=f(y)$, where $f'^2=4 f (f-1) (f-\kappa)$ we arrive at
$$
H= \frac{ d^2}{dy ^2}+n_5+n_4 (1-n_4)\,f+
\frac{n_1 (1-n_1) \kappa}{f}+\frac{n_2 (1-n_2) (1-\kappa)}{f-1}+\frac{n_3 (1-n_3) \kappa (\kappa-1)}{f-\kappa}.
$$ 
Here  $n_i$ are arbitrary parameters.
 When $n=-\frac{1}{2}(n_1+n_2+n_3+n_4)$  is a natural number, the initial polynomial operator (\ref{P11}) preserves the finite-dimensional polynomial vector space $V_n.$ 
 
Another form of this Hamiltonian (up to an additive  constant) is given by
$$
H= \frac{ d^2}{dy ^2}+n_4 (1-n_4)\,\wp(x)+
 n_1 (1-n_1) \,\wp(x+\omega_1)+
 n_2 (1-n_2) \,\wp(x+\omega_2)+
 n_3 (1-n_3) \,\wp(x+\omega_1+\omega_2),
$$ 
where $\omega_i$ are half-periods of the Weierstrass function $\wp(x)$. If $n_1=n_2=n_3=0$ we get the Lame operator. For generic case $H$ defines the so called Darboux-Treibich-Verdier model  \cite{veselov}.

\section{\bf \large Two-dimensional case}

\subsection{Structure of operators having Property 3}  The following statement describes the vector space of differential oparators we are dealing with.

{\bf Theorem 1.} An operator $L$ of the form (\ref{oper}) with polynomial coefficients satisfies Property 3 iff the coefficients of $L$ have the following structure
\begin{equation} \label{coeff}
\begin{array}{c} 
a=q_1 x^4+q_2 x^3 y+q_3 x^2 y^2+k_1 x^3+k_2 x^2 y+ k_3 x y^2+ a_1 x^2+a_2 x y+ a_3 y^2+a_4 x+a_5 y+a_6; \\[4mm]

b=q_1 x^3 y+q_2 x^2 y^2+q_3 x y^3+\frac{1}{2} \Big(k_4 x^3+(k_1+k_5) x^2 y+(k_2+k_6) x y^2+k_3 y^3 \Big)+
\\[2mm]
b_1 x^2+ b_2 x y+ b_3 y^2+b_4 x+b_5 y+b_6;
\\ \\[3mm]
c=q_1 x^2 y^2+q_2 x y^3+q_3 y^4+k_4 x^2 y+k_5 x y^2+ k_6 y^3+c_1 x^2+c_2 x y+ c_3 y^2+c_4 x+c_5 y+c_6;
 \\[3mm]
d=(1-n) \Big(2 (q_1 x^3+q_2 x^2 y+q_3 x y^2)+k_7 x^2+ (k_2+k_8-k_6) x y+k_3 y^2\Big)+d_1 x+d_2 y+d_3;
 \\[3mm]
e=(1-n) \Big(2 (q_1 x^2 y+q_2 x y^2+q_3 y^3)+k_4 x^2+(k_5+k_7-k_1) x y+
k_8 y^2\Big)+e_1 x+e_2 y+e_3;
\\[3mm]
f=n (n-1) \Big(q_1 x^2+q_2 x y+q_3 y^2+(k_7-k_1) x +(k_8-k_6) y\Big)+f_1.
\end{array}
\end{equation}

{\bf Proof.} Define a weighting scheme in the ring of differential operators with polynomial coefficients by setting weight$(x)=$weight$(y)=1$ and  weight$\frac{\pa}{\pa x}=$ weight$\frac{\pa}{\pa y}=-1$. Let
$
L=\sum_{i=-2}^m L_i
$
be the decomposition of $L$ into the sum of homogeneous differential operators $L_i$, where weight$(L_i)=i$. Let us prove that $m\le 2.$ In other words we are going to show that the degrees of polynomials $a,b,c$ in (\ref{oper}) are not greater then 4, the degrees of  $d,e$ are not greater then 3 and the degree of $f$ is not greater then 2. Let
$$
  L_m =\alpha \frac{\pa^2}{\pa x^2}+2 \beta \frac{\pa^2}{\pa x \pa y}+\gamma \frac{\pa^2}{\pa y^2}
 + \nu  \frac{\pa}{\pa x} + \mu \frac{\pa}{\pa y}+\lambda.
$$
If $m>2$ then it follows from Property 3 that $L_p(x^i y^j)=0$ for $i+j=n, n-1, n-2.$ Identities $L_p(x^n)=L_p(x^{n-1})=L_p(x^{n-2})=0$
imply $\alpha=\nu=\lambda=0$. It follows from $L_p(y^n)=L_p(y^{n-1})=L_p(y^{n-2})=0$ that $\gamma=\mu=0.$ Now the identity $L_p(y x^{n-1})=0$ leads to $\beta=0$. Thus, $m\le 2.$ It is easy to verify that the conditions  $L_2(x^i y^j)=0$ for $i+j=n, n-1$ and $L_1(x^i y^j)=0$ for $i+j=n$  are equivalent to (\ref{coeff}). The operator $L_2$ contains 3 parameters $q_i$, in $L_1$ we have 8 parameters $k_i$ 
and $L_0,L_{-1}, L_{-2}$ are arbitrary. $\square$

{\bf Remark 1.} It is easy to verify that (\ref{coeff}) is equivalent to the fact that $L$ is a quadratic  polynomial 
in generators
\[
J_1=\frac{\pa}{\pa x},\quad  J_2=\frac{\pa}{\pa y},\quad  J_3=x \frac{\pa}{\pa x},\quad  J_4=y \frac{\pa}{\pa x}, \quad  J_5=x \frac{\pa}{\pa y},\quad  J_6=y \frac{\pa}{\pa y}\ ,
\]
\begin{equation}
\label{sl3}
 J_7=x (x \frac{\pa}{\pa x}+ y \frac{\pa}{\pa y}-n), \qquad   J_8=y (x \frac{\pa}{\pa x}+ y \frac{\pa}{\pa y}-n)\ .
\end{equation}
These operators form a Lie algebra isomorphic to $sl(3).$ $\square$

Operators $J_i$ in the case of arbitrary number $k$ of independent variables have been presented in  \cite{RT:1995}. The Lie algebra generated by them is  isomorphic to $sl(k+1).$ Apparently, any polynomial operator that preserves the vector space of polynomials $V_n$ 
for sufficiently large $n$ is a polynomial in the generators $J_i.$ Following the line of proofs of Lemma 1 and Theorem 1, one can easily prove the following statement toward this conjecture.  
 
{\bf Theorem 2.} Suppose differential operator of order $m$
$$ 
L=\sum_{i_1+\cdots+i_k\le m} a_{i_1,...,i_k}  \p_{x_1}^{i_1} \cdots \p_{x_k}^{i_k} 
$$
with polynomial coefficients preserves the vector space of all polynimials in $x_1,...,x_k$ of degree $\le k,$ where $k\ge m.$ Then ${\rm deg}\, (a_{i_1,...,i_k})\le m+i_1+\cdots+i_k.$ $\square$

The dimension of the space of all operators of the form (\ref{coeff}) equals 36. The group $SL_3$ acts on this vector space as follows 
\begin{equation}\label{tran}
\tilde x=\frac{P}{R}, \qquad \tilde y=\frac{Q}{R}, \qquad \tilde L=R^{-n} L R^{n}, 
\end{equation}
where $P,Q,R$ are first degree polynomials in $x$ and $y$. The  $SL_3$ representation defined by formula (\ref{tran}) is a sum of irreducible representations with representation spaces $V_1$, $V_2$ and $V_3$ of dimensions 27, 8 and 1 correspondingly. 
The vector space $V_1$ is spanned by coefficients of the symbol. 
As a basis in $V_2$ one can choose   
$$
x_1=5 k_7-k_5-7 k_1, \qquad x_2=5 k_8-k_2-7 k_6, \qquad x_3= 5 d_1+2  (n-1) (2 a_1 +b_2),  
$$
$$
x_4= 5 e_1+2 (n-1) (2 b_1 +c_2), \quad x_5=5 d_2+2 (n-1) (2 b_3+a_2), \quad x_6=5 e_2+2 (n-1) (2 c_3+b_2),
$$
$$
x_7=5 d_3+2 (n-1) (a_4+b_5), \quad x_8=5 e_3+2 (n-1) (b_4+c_5).
$$
The generic orbit of the $SL_3$ action on $V_2$ has dimension 6. Two polynomial invariants of this action are given by 
$$
I_1=x_3^2-x_3 x_6+x_6^2+3 x_4 x_5+3 (n-1) (x_1 x_7+x_2 x_8),
$$
and
$$
I_2=2 x_3^3-3 x_3^2 x_6-3 x_3 x_6^2+2 x_6^3+9 x_4 x_5 (x_3+x_6)+
$$
$$
9 (n-1) (x_1 x_3 x_7+x_2 x_6 x_8-2 x_1 x_6 x_7-2 x_2 x_3 x_8+3 x_2 x_4 x_7+3 x_1 x_5 x_8).
$$

Assume now that an operator $L$ given by (\ref{coeff}) obeys Property 4.  Then the coefficients of $L$ have to satisfy the additional symmetry conditions
$$
\begin{array}{c}
a(x,-y)=a(x,y), \quad b(x,-y)=-b(x,y), \quad c(x,-y)=c(x,y),\\[3mm] d(x,-y)=d(x,y), \quad e(x,-y)=-e(x,y), \quad f(x,-y)=f(x,y).
\end{array}
$$
This symmetry is not destroyed by a subgroup   
\begin{equation}\label{lifr}
\tilde x=\frac{\alpha x+\beta}{\gamma x+\delta}, \qquad \tilde y=\frac{y}{\gamma x+\delta}, \qquad \alpha \delta-\beta \gamma=1.
\end{equation}
of the group (\ref{tran}) isomorphic to $SL_2.$
Transformations $\tilde L=c_1 L+c_2$ and $\tilde y=c_3 y$ are also allowed. 

Transformations (\ref{lifr}) act on 15-dimensional vector space of coefficients of polynomials $a,b,c$. The representation spaces for irreducible components of this $SL_2$-representation have dimensions $5,3,3,3,1$. Let us write the coefficients $a,b$ and $c$  in the form
\begin{equation}\label{abc}
a=P+Q y^2,\qquad b=\frac{1}{4}(P'-R) y+\frac{1}{2} Q' y^3, \qquad 
c=S+\Big(\frac{1}{12} P''-\frac{1}{4}R'+T  \Big) y^2+\frac{1}{2} Q'' y^4.
\end{equation}
where ${\rm deg}\,P=4,\quad$ ${\rm deg}\,Q={\rm deg}\,R={\rm deg}\,S=2,$ and ${\rm deg}\,T=0$. Then the coefficients of these 
polynomials  correspond to irreducible components. Namely, under transformations  (\ref{lifr}) each of these polynomials changes by the rule
\begin{equation} \label{frr}
\tilde U(x)=(\gamma x+\delta)^{{\rm deg}\,U} \, U\Big( \frac{\alpha x+\beta}{\gamma x+\delta} \Big).
\end{equation}

{\bf  Definition 1.} Symbol (\ref{abc}) of operator $L$ is called {\it elliptic} if the polynomial $P$ has four different roots on the Riemann sphere. It is called  {\it trigonometric} if $P$ has the only one double root. Sometimes we will call operators with elliptic 
(trigonometric) symbols {\it elliptic} ({\it trigonometric}). $\square$

In the elliptic case we may reduce $P(x)$ to one of the following forms:  $$P(x)=x(x-1)(x-\kappa)\qquad {\rm or} \qquad 
P(x)=(x^2-1) (x^2-\kappa)$$
by transformations (\ref{lifr}).  In the trigonometric case  we may assume  without loss of generality that 
$$P(x)=x(x-1) \qquad {\rm or\,\, that} \qquad 
P(x)=(x^2-1) x^2.$$

 {\bf Example 1.} For any constants $\alpha\ne 0,\beta, \lambda$ trigonometric operator (\ref{oper}) with coefficients
$$
a=x^2 (x^2+y^2) +\alpha x^2+\beta y^2, \quad b=x y (x^2+y^2) +(\alpha-\beta) x y,\quad c= y^2 (x^2+y^2) +\beta x^2+\alpha y^2,  $$$$ 
d=2 (n-1) x (\lambda-x^2-y^2), \qquad e=2 (n-1) y (\lambda-x^2-y^2), \qquad f=n (n-1) (x^2+y^2)
$$
satisfies Properties 1-4. The operator possesses the discrete group of symmetries isomorphic to $D_4$, generated by the reflections 
\begin{equation}\label{D4}
x \to -x,\, y \to y, \qquad  x\to x,\, y \to -y, \qquad  x\to y,\, y \to x.
\end{equation}

\subsection{Classification of flat symbols}

The contravariant metric  associated with an operator of the form (\ref{oper}) is defined by 
\begin{equation}\label{metric}
g^{1,1}=a, \qquad g^{1,2}=g^{2,1}=b,\qquad  g^{2,2}=c.
\end{equation}
This metric is flat iff $R_{1,2,1,2}=0,$ where $R_{1,2,1,2}$ is the only non-trivial component of the curvature tensor.  
This is equivalent to an identity of the form
\begin{equation}\label{curv}
\begin{array}{c}
2 \Big( b^2 a_{xx}-2 a b b_{xx}+a^2 c_{xx}+2 b c a_{xy}-2 (b^2+ac) b_{xy}+2 ab c_{xy}+\\[3mm]
c^2 a_{yy}-2 bc b_{yy}+b^2 c_{yy}\Big)\times D+{\rm first\,\, order\,\, terms} =0,
\end{array} 
\end{equation}
where we denote by $D(x,y)$ the determinant $a(x,y) c(x,y)-b(x,y)^2.$ In this paper we assume that $D\ne 0.$

{\bf Example 2.} The elliptic symbol  
\begin{equation}\label{Yakobi}
a=(x^2-1) (x^2-\kappa) +(x^2+\kappa)\,  y^2,
\qquad b=x y\, (x^2+y^2+1-2 \kappa),
\end{equation}
\[
c=(\kappa-1)(x^2-1)+(x^2+2-\kappa)\,y^2+y^4
\]
is flat. It defines a linear pencil of polynomial contravariant flat metrics \cite{dub}. 

The main observation our classification is based upon  is the following

{\bf Lemma 2.} Suppose that $a,b,c$ are given by (\ref{abc}) with $P(x)\ne 0, $  $S(x)\ne 0$ and the corresponding metric (\ref{metric}) 
is flat; then any root of the polynomial $S$ is a root of the polynomial $P.$ 

{\bf Proof.} Substituting (\ref{abc}) to (\ref{curv}) and setting $y=0,$ we get
$$
P \Big( -4 P S^2 P''+8 P S P' S'+S^2 P'^2+4 P S^2 R'-4 P R S S'+8 P^2 S S''-12 P^2 S'^2+16 Q S^3-R^2 S^2 \Big)=0
$$
It follows from here that $S$ is a divisor of $P^2 S'^2.$ If the roots of $S$ are distinct this proves the statement. If $S$ has a double root, then substituting  $S=S_1^2,$ where $S_1$ is a polynomial of first order, into above expression, we find that $S_1$ is divisor of $P^2 S_1'^2.$  $\square$

In the elliptic case without loss of generality we set $P(x)=x(x-1)(x-\kappa).$ The polynomial $P$ has the roots $0,1,\infty, \kappa$, which can be arbitrarely permuted by some transformations (\ref{frr}). 
It follows from Lemma 2 that there are two alternatives:  {\bf  \quad  A:}  $S$ has a multiple root, which is one of the roots of $P;$  and {\bf  \quad B:} $S$ has two distinct roots.  We may put  $\, S=\sigma x^2\,\,$ in  Case {\bf A} and $S=\sigma x (x-1)$ in Case {\bf B}. The constant $\sigma$ can be normalized to 1 by a scaling of $y$.

In both cases  seven unknown coefficients of the polynomials $Q,R,T$ can be easily found from the system of algebraic equations equivalent to (\ref{curv}). 

{\bf  Proposition 1.} \quad In Case {\bf  A} the condition   $R_{1,2,1,2}=0$  is equivalent to
$$
P(x)=x(x-1)(x-\kappa), \qquad S(x)=x^2, \qquad R(x)=-\frac{5}{3}(x^2-2 x+3 \kappa-2\kappa x),
$$ 
$$
Q(x)=\frac{1}{9}(x^2-x+1+\kappa^2-\kappa x-\kappa), \qquad T=0.   \quad \square
$$
 
 {\bf Proposition 2.} \quad In Case {\bf   B}  we get
 $$
P(x)=x(x-1)(x-\kappa), \qquad S(x)=x (x-1), \qquad R(x)=-3 (x^2-2\kappa x+\kappa),
$$ 
$$
Q(x)=\frac{1}{2}(x^2-2 \kappa x+2 \kappa^2-\kappa), \qquad T=\frac{1}{3}(2 \kappa-1). \quad \square
$$
It is easy to verify that the symbol  from Proposition 2 can be converted into  the symbol (\ref{Yakobi}) from Example 2 by a transformation of the form  (\ref{lifr}).

Consider now the trigonometric case $P(x)=(x-x_0)^2 (x-x_1) (x-x_2).$ According to Lemma 2 we have the following non-equivalent  possibilities:
$$
 \quad {\bf a:} \quad  S(x)=(x-x_1)^2, \qquad {\bf b:} \quad  S(x)=(x-x_1) (x-x_2), $$$$ {\bf c:} \quad  S(x)=(x-x_0) (x-x_1), \qquad {\bf d:} \quad  S(x)=(x-x_0)^2
.$$ Without loss of generality we fix $x_0=\infty, x_1=0, x_2=1$ or, in other words,  $P=x(x-1).$  

{\bf Proposition 3}. In Case {\bf a} there exists only one trigonometric symbol given by
$$
P=x(x-1), \quad S=x^2,\quad Q=\frac{1}{9}, \quad R=\frac{5}{3}(3-2 x), \quad T=0, \quad \square
$$

{\bf Proposition 4}. In Case {\bf b} we get
$$
P=x(x-1), \quad S=x (x-1),\quad Q=1, \quad R=3 (1-2 x), \quad T=-\frac{2}{3}, \quad \square
$$

{\bf Proposition 5}. In Case {\bf c} we have
$$
P=x(x-1), \quad S=x,\quad Q=\frac{1}{2}, \quad R=3, \quad T=\frac{1}{3}. \quad \square
$$

{\bf Proposition 6}. In Case {\bf d} there exist two families of trigonometric symbols depending on arbitrary parameter $\sigma.$ Up to scalings they are given by  
$$
a(x,y)=x(x-1), \qquad b(x,y)=0, \qquad   c(x,y)=1+\sigma y^2
$$
and by
$$
a(x,y)=(2 x-1)^2 y^2 + 
   \sigma (4 x^2-4 x - y^2), \quad b(x,y)=2 (2 x-1)\,y \,(1 + y^2) , \quad c(x,y)=4 (1+y^2)^2.
$$
$\square$

\subsection{Property 2 and linear terms for elliptic symbols.}   Property 2 means that we can reduce $L$ to a potential form 
$$
\bar L=\Delta_g+U
$$ 
by a gauge transform $\bar L=h L h^{-1}$ with a proper function $h.$ Here  $\Delta_g$ is the  Laplace-Beltrami operator corresponding to the metric  (\ref{metric}). It is easy to show that it is possible iff the coefficients of (\ref{oper}) satisfies the relation  
\begin{equation}\label{poten}
\frac{\partial}{\partial y} \Big( \frac{b e-c d+c(a_x+b_y)-b(b_x+c_y)}{D}   \Big)
\end{equation}
$$
=\frac{\partial}{\partial x} \Big( \frac{b d-a e+a(b_x+c_y)-b(a_x+b_y)}{D}   \Big),
$$
where $D=a c - b^2.$ Properties 1 and 2 guaranty that $L$ can be reduced to the form
$$
\bar L=\frac{\pa^2}{\pa x^2}+\frac{\pa^2}{\pa y^2}+U(x,y)
$$
 by a gauge transform and by a proper (complex) change of variables.  
 
 Notice that given polynomial symbol of operator (\ref{oper}), (\ref{coeff}), condition  (\ref{poten}) is equivalent to a system of algebraic {\bf linear} equations for the coefficients of polynomials $d$ and $e$. So, for any symbol from Section 3.2 we can easily found admissible $d$ and $e$. The coefficient $f$ is uniquely defined up to an additive constant by formulas (\ref{coeff}). 
 
For example,  for the elliptic symbol from Proposition 1 condition  (\ref{poten}) leeds to
$$
d=\frac{1}{9} (1-n)\,\Big( 3 (5 x^2-4 x-4 \kappa x+3 \kappa)+(2 x-1-\kappa)\, y^2\Big), 
$$
$$
e=\frac{2}{9} (1-n)\, y \,\Big(9 x+y^2-6 \kappa-6\Big), \qquad f=\frac{1}{9} n (n-1)\, \Big(6 x+y^2\Big). \ 
$$
Therefore, in this case we have no arbitrary constants in $d$ and $e$ except for the coupling constant $n$ and the elliptic parameter $\kappa$.
 
 However there is a non-trivial observation concerning parameters in $d$ and $e$. Due to the involution $y \to -y$ the symbol admits the transformation   $\bar x=x,\, \bar y=y^2.$ For polynimial symbol thus obtained we can again find addmissible $d$ and $e$. It turns out that now they contain an additional parameter becides  $n$ and $\kappa\,$ !  We consider the corresponding model in the next Section in details.

\section{\bf  \large $A_2$ and $G_2$ elliptic models} 

Applying the transformation $\bar x=x, \bar y=y^2$ to the symbol from Proposition 1, we get a new  polynomial symbol with 
$$
a=x (x-1)(x - \kappa) + \frac{1}{9} (1 - x + x^2 -  \kappa -  \kappa x +  \kappa^2)\,y,
$$
\begin{equation}\label{g2}
b= \frac{1}{3} ( 7 x^2 -8 x  - 8 \kappa x+9 \kappa)\,y+ \frac{1}{9} (2 x -1 - \kappa)\,y^2 ,
\end{equation}
$$
c=4 x^2 y + \frac{4}{3}  (4 x-3 - 3 \kappa)\,y^2 + \frac{4}{9}\,y^3 .
$$
The determinant $D$ for this symbol is given by
$$
D=-\frac{y}{27}\,K, \qquad  K=k_3 y^3+6 k_2 y^2+9 k_1 y+108 k_0,
$$
where 
$$
k_3=(\kappa-1)^2,\quad  k_0=x^3 (x-1)(x-\kappa), $$$$k_2=(\kappa+1) x^2+2 (\kappa^2-4\kappa +1) x-(\kappa-2)(\kappa+1)(2\kappa-1),  
$$
$$
k_1= x^4+8 (\kappa +1) x^3-2 (4 \kappa^2+23 \kappa +4) x^2+36 \kappa (\kappa+1)x-27 \kappa^2. 
$$
The transformation  $\bar x=x,\, \bar y=y^2$ changes the weighting scheme in the ring of differential operators. For the new variables we have  weight$(x)=1$, weight$(y)=2$, weight$\frac{\pa}{\pa x}=-1$ and  weight$\frac{\pa}{\pa y}=-2$.  Denote  by $W_n$ the vector space spanned by monomials $x^i y^j,$ where $i+2 j\le n.$

{\bf Theorem 3.} An operator $L$ of the form (\ref{oper}) with polynomial coefficients preserves $W_n,$ $n\ge 4,$ iff the coefficients of $L$ have the following structure

\begin{equation} \label{coeff12}
\begin{array}{c} 
a= k_1 x^4 +k_2 x^2 y +k_3 x^3 + k_4 x y +  a_1 y + a_2 x^2 + a_3 x + a_4; \\[4mm]

b=2 k_1 x^3 y + 2 k_2 x y^2 + k_4 y^2 +k_5 x^4 + k_6 x^2 y + 
b_1 x y + b_2 x^3 + b_3 y + b_4 x^2 + b_5 x + b_6;
 \\[4mm]
c=4 k_1 x^2 y^2 + 
    4 k_2 y^3 + 4 k_5 x^3 y - 
    4 (k_3 - k_6) x y^2 + c_1 y^2 + c_2 y x^2 + \\[2mm]
    c_3 x^4 + c_4 x y +  c_5 x^3 + c_6 y + 
    c_7 x^2 + c_8 x + c_9;
 \\[3mm]
d=(1-n)\,  \Big( 2 k_1 x^3 + 2 k_2 x y + k_4 y \Big) + k_7 x^2 +d_1 x + d_2;
 \\[3mm]
e=2 (3 - 2 n)\,  \Big(k_1 x^2 y + k_2  y^2 - k_3 x y\Big) + 
 2 (2 - n)  \Big(k_5 x^3 + k_6 x y\Big) + 2 k_7 x y+e_1 y + e_2 x^2 + e_3 x + e_4;
\\[3mm]
f=(n - 1) n\,  \Big(k_1 x^2 + k_2 y - k_3 x\Big) - n\,k_7 x + f_1. \qquad \square
\end{array}
\end{equation}
We will skip the "spectral" parameter $f_1$ in all futher formulas.

Writing $L$ in the form (\ref{coeff12}), where $a,b$ and $c$ are defined by (\ref{g2}), we find that
condition (\ref{poten}) implies 
$$
d=\frac{1}{9} (1 - n) (6 x + y) (2 x-1 - \kappa) + \frac{s}{3} (x^2   - 2 x -2 \kappa x+3 \kappa),
$$
 $$
e= \frac{2}{9} (9 x^2 + 12 x y + y^2 - 9 y  - 9 \kappa y) + \frac{2 s}{3} (3 x^2 +x y- y- \kappa y) + $$$$ \frac{2  (n-1) }{9}
     (9 x^2 - 15 x y - 2 y^2  + 9 y+  9 \kappa y),
$$
$$
f= \frac{n(n-1)}{9} (3 x + y)-\frac{s}{3} n x. 
$$
We see that this formulas involve an extra parameter $s$. 

\medskip 

\qquad Introduce parameters $m_i$ by identities 
$$
n=-3 m_1-m_2, \qquad s=1+3 m_1+3 m_2.
$$
  Then 
\begin{equation}\label{lapbel}
h L h^{-1}=\Delta_g+m_2(1-m_2)\frac{x^2}{y}+3 m_1(1-m_1)\frac{P^2}{K}+\lambda.
\end{equation}
Here $\Delta_g$ is the Laplace-Beltrami operator,  $\quad h=K^{\frac{m_{1}}{2}} y^{\frac{m_{2}}{2}},$ and
$$
P=3 x^3-6(\kappa+1)\, x^2+(y+\kappa y+9 \kappa)\, x-2 (\kappa^2-\kappa+1) y,
$$
$$
\lambda=\frac{\kappa+1}{3}(3 m_1+m_2)(1+3 m_1+3 m_2).
$$

Now we are to find a transformation of the form  $x=\phi(y_1,y_2), \, y=\psi(y_1,y_2)^2$ that brings $\Delta_g$ to the form (cf.  (\ref{HAM}))
$$
\Delta= -\frac{1}{3}\left(\frac{\partial^2}{\partial y_1^2}+\frac{\partial^2}{\partial y_2^2}-\frac{\partial^2}{\partial y_1 \partial y_2} \right). 
$$
Loking for a transformation of the form of formal series
$$
\phi=\sum_{i=0}^{\infty} \phi_i(y_2) y_1^i, \qquad \psi=\sum_{i=1}^{\infty} \psi_i(y_2) y_1^i, 
$$
we find several first coefficients of these series in terms of $\phi_0$ and its derivatives, derive a first order ODE for $\phi_0,$ and 
after all write the transformation in the following closed form 
 $$
 x=\frac{f(y_1)^2 f'(y_2)-f(y_1)^2 f'(y_2)}{f(y_1) f'(y_2)-f(y_1) f'(y_2)}, \qquad 
y=-12 \left(\frac{f(y_1) f(y_2) (f(y_1)- f(y_2))}{f(y_1) f'(y_2)-f(y_1) f'(y_2)}\right)^2,
 $$
 where $f'^2=4 f (f-1) (\kappa-f).$ Under this transformation the operator (\ref{lapbel}) becomes
$$
 {\cal H}= -\frac{1}{3}\left(\frac{\partial^2}{\partial y_1^2}+\frac{\partial^2}{\partial y_2^2}-\frac{\partial^2}{\partial y_1 \partial y_2} \right)+V(y_1,y_2),
$$
where
$$
V=(m_1-1)m_1 \Big(\wp(y_1-y_2)+\wp(2 y_1+y_2)+ \wp(y_1+2 y_2) \Big)+
$$
$$
+ \frac{(m_2-1)m_2}{3} \Big(\wp(y_1)+\wp(y_2)+ \wp(y_1+y_2) \Big).
$$
This is just the elliptic $G_2$ Calogero-Moser model \cite{perel}. The elliptic $A_2$-model corresponds to the special case $m_2=0.$
The invariants of the $\wp$-function are related to the parameter $\kappa$ as follows
$$
g_2=\frac{4}{3}(\kappa^2-\kappa+1), \qquad g_3=-\frac{4}{27}(\kappa-2)(\kappa+1)(2\kappa-1).
$$
The polynomial form of the $G_2$-model  preserves $W_n$ if
$
n=-3 m_1-m_2 \quad
$
is a natural number.

\section{\bf \large   The Inozemtsev elliptic model} 
 
In this section we investigate the model related to the symbol from Proposition 2 written in the form  (\ref{Yakobi}).
 Polynomials $d$ and $e$ for symbol  (\ref{Yakobi}) depend on three arbitrary parameters becides $n$ and $\kappa$. But due to $D_4$-symmetry the symbol admits the transformation   $\bar x=x^2,\, \bar y=y^2.$ As a result of this transformation 
and a scaling we get
\[
a=x (x-1) (x-\kappa) +(1-\kappa)\,x (x+\kappa)\, y,
\]
\[
 b=x (x+1-2\kappa)\,y +(1-\kappa)\,x \,  y^2,
\]
\[
c=(1-x)\,y+ (x+2-\kappa)  \,y^2+(1-\kappa)\,y^3.
\]
For this symbol we still may use the anzats of Theorem 1. 

 It follows from (\ref{poten}) that 
\[
d=\lambda_1 x (x+y-\kappa y) +\lambda_2 (1+y-\kappa y)+ p\,x,
\]
\[
e=\lambda_1 y (x+y-\kappa y) +\lambda_3 (x-1)+ q\,y,
\]
\[
f=\lambda_4 (x+y-\kappa y)+\lambda_5,
\]
where 
\[
\kappa p+(1-\kappa) q+\lambda_1 (2 \kappa-1) +\lambda_2 (2-\kappa)+ \lambda_3(1-\kappa^2)=0.
\]
Thus,  $d$ and $e$ depend on five arbitrary parameters ! Define constants $m, n_0,n_1,$ $n_2, n_3$ from relations  
$$
2 \lambda_1 =  3 + 4 m + 2 n_1 + 
         2 n_2 + 2 n_3,\qquad   2 \lambda_2 = 
     \kappa (1 + 2 n_2), \qquad 2 \lambda_3 = -1 - 2 n_1, $$$$ -2 \lambda_4 = n_0 (1 + 4 m + 
         2 n_0 + 2n_1 + 2 n_2 + 2 n_3),
$$
$$
2 p = -1 - 3 \kappa - 4 \kappa m + 
         2 n_1 - 4 \kappa n_1 - 2 n_2 - 
         2 \kappa n_2 - 2 n_3, $$$$
2  q = 4 - 3 \kappa + 4 m - 
         4 \kappa m + 4 n_1 - 2 \kappa n_1 + 
         2 n_2 - 4 \kappa n_2 + 2 n_3. 
$$
 
Let us put $\bar L=- 4 L,$ where $L$ is the operator  (\ref{oper}) with the polynomial coefficients defined above. 
Reducing the operator $L$ to the Schrodinger form by the transformation 
$$
\bar x=f(x) f(y), \qquad \bar y=\frac{(f(x)-1) (f(y)-1)}{\kappa-1},
$$
where
$$
 f'^2=4 f (f-1) (f-\kappa),
$$
and by a proper gauge transformation, we get 
$$
H=\Delta+ 2 m (m-1) \Big(\wp(x+y)+\wp(x-y)\Big)+\sum_{i=0}^3 n_i(n_i-1) (\wp(x+\omega_i)+\wp(y+\omega_i)),
$$
where  $\omega_1,\omega_2$ are the half-periods of 
the Weierstrass function $\wp(x)$, $\omega_0=0$, and $\omega_3=\omega_1+\omega_2$. 

This is so called Inozemtsev $BC_2$ Hamiltonian \cite{Inoz}.  Its polynomial form   preserves $V_k$ if
$$
k=-\frac{1}{2} (2 m+\sum n_i)
$$
is a natural number.

\end{document}